\documentclass{JHEP3}

\newcommand{\hs}{\hspace{0.15mm}}

\usepackage{bm}
\usepackage{latexsym}
\usepackage[latin1]{inputenc}
\usepackage{amsfonts}
\usepackage{amssymb}
\usepackage{amsmath}

\numberwithin{equation}{section}

\title{No chiral truncation of quantum log gravity?}
\author{Tom\'{a}s Andrade, Donald Marolf \\  Department of Physics, UCSB, Santa Barbara, CA 93106, USA }

\abstract{At the classical level, chiral gravity may be constructed as a consistent truncation of a larger theory called log gravity by requiring that left-moving charges vanish.  In turn, log gravity is the limit of topologically massive gravity (TMG) at a special value of the coupling (the chiral point).  We study the situation at the level of linearized quantum fields, focussing on a unitary quantization.  While the TMG Hilbert space is continuous at the chiral point, the left-moving Virasoro generators become ill-defined and cannot be used to define a chiral truncation.  In a sense, the left-moving asymptotic symmetries are spontaneously broken at the chiral point.    In contrast, in a non-unitary quantization of TMG, both the Hilbert space and charges are continuous at the chiral point and define a unitary theory of chiral gravity at the linearized level. }

\date{\today}


\begin{document}

\section{Introduction}
\label{section introduction}

It was recently proposed \cite{chiral 1 strominger} that a 2+1 dimensional theory known as chiral gravity could be defined at the quantum level, providing an interesting yet perhaps exactly solvable model of both quantum gravity and the anti-de Sitter/Conformal Field Theory (AdS/CFT) correspondence \cite{AdS/CFT maldacena,AdS/CFT gubser klebanov polykov, AdS/CFT witten}.   In particular, it was suggested that chiral gravity is dual to an extremal chiral CFT, and further supporting evidence was provided in \cite{chiral & log gravity and extremal cft strominger}.  This situation contrasts with that of pure gravity, where an exact CFT dual has not yet been understood \cite{witten 3d gravity solvable,Witten partition function,witten 3d gravity reconsidered}.

On the other hand,  extremal CFT's with large central charges have not been constructed, and it has been argued that they do not exist \cite{gaberdiel no extremal cft}. Our goal is to investigate this tension by taking a brief first look at the quantum theory.  Because chiral gravity boundary conditions remove the local propagating degree of freedom that would otherwise arise, one might be suspicious that the dynamics of this theory are ill-defined.  However, at the classical level chiral gravity may be defined as the truncation of a larger theory, called log gravity, to the sector defined by requiring certain charges to vanish \cite{chiral & log gravity and extremal cft strominger}.   Conservation of the charges then implies consistency of this truncation.  Log gravity is described by the same action, but with boundary conditions that allow the expected local degree of freedom.  The asymptotic symmetry group of log gravity contains two Virasoro algebras, and one arrives at chiral gravity when the left-moving charges vanish.   Log gravity may itself be defined as the limit of topologically massive gravity (TMG) at a special value of the coupling known as the chiral point.

Below, we investigate the situation at the level of linearized quantum fields.  As we briefly discuss in section \ref{NU}, a non-unitary quantization of linearized log gravity reproduces the classical story and leads to a unitary theory of chiral gravity.  However, this approach can succeed at higher orders in perturbation theory only if certain ghost-modes continue to decouple in an appropriate way.  In addition, since there appears to be a sensible (if unstable) classical theory of log gravity, one would expect the physics of log gravity to be better captured by a unitary quantization of the the log gravity theory.

For these reasons we focus on a unitary quantization of log gravity below.  We construct the quantum theory using a unitary quantization of TMG away from the chiral point and taking an appropriate limit.  While the Hilbert space and right-moving charges are continuous at the chiral point, the left-moving charges become ill-defined.  In a sense, the left-moving symmetries are spontaneously broken at the chiral point.  As a result,  they cannot be used to define a chiral truncation.

After reviewing the classical theory of anti-de Sitter topologically massive gravity and computing the symplectic structure in section \ref{prelim}, we discuss the unitary quantum theory in section \ref{newQ}.  This section shows that the Hilbert space defined by our unitary quantization of TMG is continuous at the chiral point. The quantum charges are studied in section \ref{CandC} for both unitary and non-unitary quantizations. We close with some discussion of open questions in section \ref{disc}.

\section{Preliminaries and Notation}
\label{prelim}

Chiral gravity is a special case of  Topologically Massive Gravity (TMG) with negative cosmological constant (TMG) \cite{TMG birth 1,TMG birth 2,CTMG deser} defined by a certain relation between  coupling constants and a particular choice of boundary conditions.    We begin by reviewing this basic setting.  As noted in the introduction, we will benefit from a unified perspective taking into account all values of the coupling.

The TMG action is
\begin{equation}\label{action}
    I = \frac{1}{16 \pi G} \left[ \int d^3x \sqrt{-g}(R - 2\Lambda) +   I_{CS}  \right],
\end{equation}
where $I_{CS}$ is the gravitational Chern-Simons term
\begin{equation}\label{ICS}
    I_{CS} = \frac{1}{2\mu} \int d^3x \sqrt{-g} \epsilon^{\alpha \beta \gamma} \Gamma^{\rho}_{\alpha \sigma}[ \partial_\beta \Gamma^{\sigma}_{\rho \gamma}
    + \frac{2}{3} \Gamma^\sigma_{\beta \lambda} \Gamma^\lambda_{\gamma \sigma} ].
\end{equation}
Our conventions for the curvature and the Levi-Civita symbol are $R^\alpha \hs_{\beta \mu \nu} = \partial_\mu \Gamma^\alpha_{\beta \nu} + \ldots$ and $\epsilon^{\rho t \phi} = +1$ respectively. The theory is power counting renormalizable \cite{TMG renormalizable oda}.

Since (\ref{ICS}) is parity odd, we may choose $\mu > 0$ without loss of generality. We are interested in linear perturbations around AdS${}_3$, whose line element in global coordinates reads:
\begin{equation}\label{AdS}
    ds^2 = \bar{g}_{\mu \nu} dx^\mu dx^\nu = l^2 (d \rho^2 - \cosh^2 \rho dt^2 + \sinh^2 \rho d \phi^2).
\end{equation}
The metric  (\ref{AdS})  is a stationary point of (\ref{action}) for $\Lambda = - 1/ l^2$.  From now on, we take $l=1$ unless otherwise specified.

The discussion for $\mu = 1$ is somewhat subtle, so we first consider $\mu  \neq 1$.  In this case one imposes the Brown-Henneaux  boundary conditions (BHBCs) described in \cite{brown henneaux} for pure Einstein-Hilbert gravity.  BHBCs admit an asymptotic symmetry group generated by the vector fields
\begin{equation}\label{Killing assoc to Ln}
    \xi_n = i e^{in(t+\phi)}\left\{ -\frac{in}{2} \partial_\rho + [\frac{1}{2} - n^2e^{-2\rho}]\partial_t  +
    [\frac{1}{2} + n^2e^{-2\rho}]\partial_\phi \right \} + \xi_{gauge},
\end{equation}
\begin{equation}\label{Killing assoc to barLn}
    \bar{\xi}_n = i e^{in(t-\phi)}\left\{ -\frac{in}{2} \partial_\rho + [\frac{1}{2} - n^2e^{-2\rho}]\partial_t  -
    [\frac{1}{2} + n^2e^{-2\rho}]\partial_\phi \right \} + \xi_{gauge},
\end{equation}
where $\xi_{gauge}$ falls off rapidly enough at infinity that it defines a gauge transformation\footnote{By gauge transformation, we mean a degenerate direction of the symplectic structure; see section \ref{section quantization}.}; i.e.
\begin{equation}\label{xi pure gauge}
    \xi_{gauge} = {\cal O}(e^{-8\rho})\partial_\rho + {\cal O}(e^{-4\rho}) \partial_t +  {\cal O}(e^{-4\rho}) \partial_\phi \,\,\,\, .
\end{equation}
We will refer to (\ref{Killing assoc to Ln}) and (\ref{Killing assoc to barLn}) as left and right symmetries henceforth. These vector fields satisfy the Witt algebra
\begin{equation}\label{witt algebra}
    [\xi_n, \xi_m] = (n-m)\xi_{n+m} \,\,\,\,\,\,\,\,\,\,\,\,\, [\bar{\xi}_n, \bar{\xi}_m] = (n-m)\bar{\xi}_{n+m}.
\end{equation}
For $n=0, \pm1$, the vector fields (\ref{Killing assoc to Ln}) and (\ref{Killing assoc to barLn}) generate the $SL(2,R)_L \times SL(2,R)_R$ isometry group of AdS${}_3$. Below, we use the notation $L_i, \bar L_j$ for Lie derivatives along $\xi_i, \bar \xi_j$, where the terms $\xi_{gauge}$ are chosen to make all $\xi_i$ smooth.  The charges associated with (\ref{Killing assoc to Ln}) and (\ref{Killing assoc to barLn}) (which we also call $L_n$ and $\bar L_n$) satisfy the Virasoro central extension of (\ref{witt algebra}) with central charges \cite{solodhukin HR TMG,larsen krauss}
\begin{equation}\label{central charges generic mu}
    c_L = \frac{3}{2G}(1-\frac{1}{\mu})  \,\,\,\,\,\,\,\,\, c_R = \frac{3}{2G}(1+\frac{1}{\mu}).
\end{equation}

We now review the linearized modes following \cite{chiral 1 strominger, grumiller log}.  These modes may be classified using the  $SL(2,R)_L \times SL(2,R)_R$ symmetry of the background.  The three $SL(2,R)$ primaries with their left and right conformal weights are:
\begin{equation}\label{primaries generic mu and weights}
    \Psi_L \,\, (2,0) \,\,\,\,\,\,\,\, \Psi_R \,\, (0,2) \,\,\,\,\,\,\,\, \Psi_M \,\, (\frac{3}{2} + \frac{\mu}{2}, -\frac{1}{2} + \frac{\mu}{2}).
\end{equation}
The explicit wave functions can be found in \cite{chiral 1 strominger}. It suffices for our purposes here to say that their $t$ and $\phi$ dependence occurs only through complex exponentials. The descendants are obtained by acting on (\ref{primaries generic mu and weights}) with $L_{-1}$ and $\bar{L}_{-1}$. Thus the modes of this theory are uniquely specified by three labels:
\begin{equation}\label{definition descendants}
    \Psi^{\alpha \bar{\alpha}}_{A} = L_{-1}^\alpha \bar{L}_{-1}^{\bar {\alpha} } \Psi_A,
\end{equation}
\noindent where the index $A$ runs over the three primaries,  $A\in \{ L,R,M \}$. A complete set of solutions obeying BHBCs consists of (\ref{primaries generic mu and weights}), their descendants and complex conjugates. This is consistent with the analysis made independently in \cite{carlip deser photons gravitons} where it is shown that the only propagating degree of freedom corresponds to a single scalar. For later use, we record the fact that the Virasoro descendants of AdS${}_3$ take the form
\begin{equation}\label{Lminusa bar g prop Psi L}
    L_{-\alpha} \bar{g} \approx \frac{3}{(\alpha-2)!} \Psi_L^{(\alpha-2)}, \ \ \
    \bar{L}_{-\bar{\alpha}} \bar{g} \approx \frac{3}{(\bar{\alpha}-2)!} \Psi_R^{(\bar{\alpha}-2)},
\end{equation}
where $\bar g$ represents AdS${}_3$ and $\approx$ indicates equality up to pure gauge modes.
This fact follows by direct calculation for $\alpha =2$ and thence from the algebraic relation $L_{-\alpha} = \frac{1}{\alpha-2}[L_{-1},L_{-(\alpha-1)}]$ or the equivalent for the right-moving charges.

One sees from (\ref{central charges generic mu}) that the limit $\mu \to 1$ is special since $c_L \to 0$.  In addition, it turns out \cite{chiral 1 strominger} that $\Psi_L - \Psi_M \to 0$ as $\mu \to 1$, as suggested by the fact that their conformal weights (\ref{primaries generic mu and weights}) coincide in this limit. The basis given by (\ref{primaries generic mu and weights}) and their descendants must therefore be supplemented  \cite{grumiller log} by another linearly independent mode:
\begin{equation}\label{psi log}
\Psi_{log} \equiv \mathop {\lim }\limits_{\mu \to 1 } \frac{ \Psi_M(\mu) - \Psi_L}{{\mu-1} }= (-it -log(\cosh \rho)) \Psi_L.
\end{equation}
As emphasized in \cite{grumiller log}, $\Psi_{log}$ has a qualitatively different behavior than that of the primaries, since its time dependence is not exponential and it grows as $\log \rho$ for large $\rho$. Non-linear configurations exhibiting such `logarithmic' behavior at the chiral point were found previously in \cite{AyonBeato:2004fq, AyonBeato:2005qq}. This means that $\Psi_{log}$ does not satisfy BHBCs.  The $\mu =1$ theory with BHBCs (and thus without the mode $\Psi_{log}$) is called chiral gravity.

However, one can consistently relax BHBCs  to so-called log boundary conditions to accommodate $\Psi_{log}$ \cite{valdivia AdS spaces in TMG, Grumiller consistent bc, chiral & log gravity and extremal cft strominger}.  The log boundary conditions  again lead to classical charges generating two copies of the Virasoro algebra, which are just the $\mu \to 1$ limit of those for $\mu \neq 1$.    The resulting theory is called log gravity.  It turns out that $\Psi_{log}$ is not an eigenstate of either $L_0$ or $\bar{L}_0$, and so is not strictly-speaking a primary. As noted in \cite{grumiller log},
$\Psi_{log}$ is properly referred to as a ``log-primary'' in the language of log-CFT's \cite{LCFT log operators in CFT gurarie}, see \cite{LCFT bits and pieces flohr} and \cite{LCFT algebraic approach gaberdiel} for reviews.   For $\mu=1$, a general solution obeying the log-boundary conditions consists of an arbitrary linear combination of $\Psi_L$, $\Psi_R$, $\Psi_{log}$ and their $SL(2,R)$ descendants\footnote{The descendants $\Psi_{log}^{\alpha \bar{\alpha}}$ of $\Psi_{log}$ are slightly subtle.
The logarithmic tail of $\Psi_{log}^{\alpha \bar{\alpha}}$ for $(\alpha , \, \bar{\alpha}) \neq (0,0)$ can be removed by a gauge transformation at the linearized level  \cite{GKP}. However, this is no longer true at second order in perturbation theory, where they violate BHBCs \cite{chiral & log gravity and extremal cft strominger}.}.   See \cite{carlip chiral TMG and extremal BF} for an independent analysis of the propagating degrees of freedom at the chiral point and \cite{HR for TMG skenderis taylor} for more on the relationship between log gravity and log CFT's.

As discussed in \cite{simple proof strominger,chiral & log gravity and extremal cft strominger}, chiral gravity may also be defined as the truncation of log gravity to the sector in which the left-moving charges $L_{n}$  vanish.  While the log gravity Hamiltonian is unbounded below, it has been argued \cite{chiral & log gravity and extremal cft strominger} that the constraints $L_{n} =0$ render this Hamilton positive definite.  This was definitively established at the linearized level at which we work here.

Because chiral gravity boundary conditions remove the local propagating degree of freedom that would otherwise arise, one might a priori be suspicious that the dynamics of this theory are ill-defined.  Defining chiral gravity as the above truncation of log gravity removes this concern at the classical level.   It is therefore of interest to learn whether a similar truncation is possible at the quantum level.
At least for a natural unitary quantization of log gravity, we show in section \ref{CandC} that the quantum construction fails in the  linearized approximation.  On the other hand, it succeeds in this approximation for a non-unitary quantization.

\subsection{The symplectic structure}
\label{section quantization}

Our goal is to examine the quantum constraints defined by the left-moving charges $L_n$ at the chiral point.  To do so, we must first  quantize the theory.  This will be done in section \ref{newQ}
 below, where we use an operator method based on the covariant phase space formalism.  An important ingredient will be the symplectic structure, which we now compute.

The symplectic structure is defined as follows.   Given a Lagrangian density $L(\phi)$, where $\phi$ denotes an arbitrary collection of fields, we consider a small deformation $\delta_1 \phi$ away from a background configuration $\bar{\phi}$. This variation can always be written $\delta_1 L \approx \nabla_\mu \theta^\mu(\delta_1 \phi, \bar{\phi})$, where $\approx$ denotes equivalence on-shell. We now consider another independent variation $\delta_2 \phi$ and define the symplectic current
\begin{equation}\label{definition omega}
   \omega^\mu(\delta_1 \phi, \delta_2 \phi; \bar{\phi}) = \frac{1}{\sqrt{-g}} \left( \delta_2 \theta^\mu(\delta_1 \phi, \bar{\phi}) - \delta_1 \theta^\mu(\delta_2 \phi, \bar{\phi}) \right) ,
\end{equation}
which is conserved $\nabla_\mu \omega^\mu = 0$ when $\bar{\phi} + \delta_{1} \phi$ and  $\bar{\phi} + \delta_{2} \phi$  solve the equations of motion.
Furthermore, if the symplectic flux through the boundary vanishes by the boundary conditions, then the integral
 \begin{equation}\label{definition OMEGA}
    \Omega(\delta_1 \phi, \delta_2 \phi; \bar{\phi}) = \int_{\Sigma} n_\mu \omega^\mu(\delta_1 \phi, \delta_2 \phi; \bar{\phi})
\end{equation}
\noindent is independent of the choice of the space-like surface $\Sigma$. In (\ref{definition omega}), the integral uses the volume measure on $\Sigma$.  We refer to $\Omega$  as the {\it symplectic structure} of the theory. See \cite{wald lee symplectic} for a detailed construction.

Given two complex linearized solutions $\varphi_{1}$, $\varphi_{2}$, it is convenient to replace $\Omega$ by the Hermitian (but not positive definite) inner product
\begin{equation}\label{inner product}
    (\varphi_1, \varphi_2) =  -i \ \Omega(\varphi_1, \varphi_2^*).
\end{equation}
This product allows us to define some useful terminology.  We use the convention that a positive frequency mode is a normal particle/ghost if its symplectic norm is positive/negative, with the opposite convention for negative frequency modes.   In addition, a mode is pure gauge if it has vanishing symplectic product with  all modes, so that it defines a degenerate direction of $\Omega$.  After moding out by these null directions, one may invert the symplectic structure to define a Poisson bracket for gauge invariant functions on the space of solutions.

For the action (\ref{action}) around the AdS background, we find the symplectic current,
\begin{equation}\label{omega TMG}
    \omega^\nu_{TMG} = \omega^\nu_{EH}  + \omega^\nu_{CS},
\end{equation}
\noindent where
\begin{equation}\label{w EH}
    \omega^\nu_{EH} = \frac{1}{16 \pi G} \bigg [ \delta_2(\sqrt{g}g^{\alpha \beta} ) \delta_1 \Gamma^\nu_{\alpha \beta} - \delta_2(\sqrt{g}g^{\nu \alpha} ) \delta_1 \Gamma^\beta_{\alpha \beta} \bigg] - (1 \leftrightarrow 2)
\end{equation}
\begin{equation}\label{w CS}
     \omega^\nu_{CS} = \frac{1}{32 \pi G \mu}\epsilon^{\nu \alpha \beta}[ \delta_1 \Gamma^\lambda_{\sigma \alpha} \delta_2 \Gamma^\sigma_{\lambda \beta} - 2 \delta_1g_{\alpha \sigma} {\cal G}^{(1)}(\delta_2 g)^\sigma \hs _\beta ] -  (1 \leftrightarrow 2),
\end{equation}
and where as in \cite{chiral & log gravity and extremal cft strominger} the symbol ${\cal G}^{(1)}{}^\sigma {}_\beta$ denotes the linearization of the tensor $G^\sigma_\beta - \frac{1}{\ell^2} g^\sigma_\beta$ describing the equation of motion for $\mu = \infty$.  When evaluated around the pure AdS${}_3$ background (\ref{AdS}), the symplectic structure  leads to the norms:
\begin{equation}\label{norms primaries generic mu}
  (\Psi_L, \Psi_L) = \frac{1}{8 G} \frac{2(\mu-1)}{3\mu} \,\,\,\,\,\,\,
  (\Psi_R, \Psi_R) = \frac{1}{8 G} \frac{2(\mu+1)}{3\mu} \,\,\,\,\,\,\,
(\Psi_M, \Psi_M) =  \frac{1}{8 G} \frac{1-\mu^2}{\mu(\mu+2)},
\end{equation}
where we remind the reader that we assume $\mu > 0$.
In addition, all cross terms vanish: $(\Psi_A, \Psi_B)$ = 0 for $A \neq B$.   Note that $\Psi_L$ is a ghost for $\mu < 1$ while $\Psi_M$ is a ghost for $\mu > 1$, though otherwise we have normal (non-ghost) particles.  As usual, the ghosts carry negative energy and lead to perturbative instabilities.  For the special case $\mu=1$ the norm of $\Psi_L = \Psi_M$ vanishes.  Below, it will often  be convenient to focus on the non-degenerate case $\mu > 1$. Similar results hold for $\mu < 1$ with different choices of signs, and we will carefully take the limit $\mu \to 1$.  We stress that all modes $\Psi_L, \Psi_R, \Psi_M$ are normalizable with respect to (\ref{omega TMG}) for all $\mu$ without the addition of further boundary terms.

In computing the inner products of descendents, it is useful to note that the inner product  is invariant under AdS${}_3$ isometries in the sense that
\begin{equation}
\label{isom}
(\varphi_1, L_i \varphi_2) = ( L_{-i} \varphi_1, \varphi_2)
\end{equation}
for $i = -1,0,1$ and similarly for $\bar L_i$.  The various signs in (\ref{isom}) follow from our conventions for the vector fields (\ref{Killing assoc to Ln}), (\ref{Killing assoc to barLn}).  One may use (\ref{norms primaries generic mu}) and (\ref{isom}) to show that the descendants have inner products
\begin{equation}\label{inner prod descendants gen mu}
    (\Psi_A^{\alpha \bar{\alpha}},\Psi_B^{\beta \bar{\beta}}) = \delta^{\alpha \beta} \delta^{\bar{\alpha} \bar{\beta}} \alpha ! \bar{\alpha}! P(2h_A,\alpha)P(2\bar{h}_A,\bar{\alpha}) (\Psi_A, \Psi_B),
\end{equation}
where $(h_A, \bar{h}_A)$ are the conformal weights of the primary $\Psi_A$ and $P(a,b)$ is the Pochhammer symbol, defined through
$    P(a,b)= a(a+1)(a+2)\ldots(a+b-1)$ for $b \in \mathbb{Z}$.
Note that in particular $P(a,0)=1$, $P(0,b)=0$.   Using (\ref{inner prod descendants gen mu}), it is not hard to see that the descendants $\Psi_L^{\alpha \bar{\alpha}}$ and $\Psi_R^{\beta \bar{\beta}}$ are pure gauge for $\bar{\alpha} > 0$ and $\beta > 0$ respectively. As a result, the physical modes of TMG for $\mu >1$ are:
\begin{equation}\label{list of modes and descendants gen mu}
    \Psi_L^{\alpha} \,\, , \,\,\,\,\,\,  \Psi_R^{\bar{\alpha}} \,\, , \,\,\,\,\,\,\,  \Psi_{M}^{\alpha} \,\, , \,\,\,\,\,\,\, \Psi_{M}^{\alpha \bar{\alpha}},
\end{equation}
where we have defined $ \Psi_L^{\alpha} \equiv \Psi_L^{\alpha 0} $, $ \Psi_R^{\bar{\alpha}} \equiv \Psi_R^{0 \bar{\alpha}} $ and $ \Psi_M^{\alpha} \equiv \Psi_M^{\alpha 0}$, and  we have separated $ \Psi_M^{\alpha \bar{\alpha}}$ for $\bar{\alpha} = 0$ and $\bar{\alpha} > 0$  for future convenience. In our notation, the presence of an explicit $\bar{\alpha}$ in $ \Psi_M^{\alpha \bar{\alpha}}$ indicates that $\bar{\alpha} >0$ unless otherwise noted (though the same is not true for $\alpha$). As mentioned above, both $\Psi_L^{\alpha}$ and $\Psi_R^{\bar{\alpha}}$ define normal particles while $\Psi_{M}^{\alpha}$ and $\Psi_{M}^{\alpha \bar{\alpha}}$ are ghosts.   It is therefore convenient to define normalized fields
\begin{eqnarray}
\label{NormalizedGeneric}
    &\tilde{\Psi}_L^\alpha(\mu) =  N^{-1}_L(\mu, \alpha) \Psi_L^\alpha \,\,\,\,\, \tilde{\Psi}_R^{\bar{\alpha}}(\mu) =  N^{-1}_R(\mu, \bar{\alpha}) \Psi_R^{\bar{\alpha}}& \,\,\,\,\, \cr &\tilde{\Psi}_M^\alpha(\mu) = N^{-1}_M(\mu, \alpha)\Psi_M^\alpha \,\,\,\,\,
    \tilde{\Psi}_M^{\alpha \bar{\alpha}}(\mu) = N^{-1}_M(\mu, \alpha, \bar{\alpha}) \Psi_M^{\alpha \bar{\alpha}}&
    \end{eqnarray}
where
    \begin{eqnarray}
       &N^2_L(\mu, \alpha) = (\Psi_L^\alpha,\Psi_L^\alpha) \,\,\,\,\,\, N^2_R(\mu, \bar{\alpha}) = (\Psi_R^{\bar{\alpha}}, \Psi_R^{\bar{\alpha}})& \,\,\,\,\, \cr &N^2_M(\mu, \alpha) = -  (\Psi_M^{\alpha}, \Psi_M^\alpha) \,\,\,\,\, N^2_M(\mu, \alpha, \bar{\alpha}) = -  (\Psi_M^{\alpha \bar{\alpha}}, \Psi_M^{\alpha \bar{\alpha}}).&
\label{norms generic mu}
\end{eqnarray}
Hence the fields decorated with a tilde have norm $+1$ if they are descendants of $\Psi_{L,R}$ and $-1$ if they are descendants of $\Psi_M$, with all cross products equal to zero.

In much the same way, for $\mu =1$ we may take the physical modes of log gravity to be
\begin{equation}\label{normalized modes mu = 1}
    \hat{\Psi}_L^\alpha = N^{-1}_L(\alpha) \Psi_L^\alpha \,\,\,\,\,\,\,\, \hat{\Psi}_R^{\bar{\alpha}} = N^{-1}_R(\bar{\alpha}) \Psi_R^{\bar{\alpha}} \,\,\,\,\,\,\,\, \hat{\Psi}_{log}^\alpha = N^{-1}_n(\alpha) \Psi_{log}^\alpha \,\,\,\,\,\,\,\,
    \hat{\Psi}_{log}^{\alpha \bar{\alpha}} = N^{-1}_{GKP}(\alpha, \bar{\alpha}) \Psi_{log}^{\alpha \bar{\alpha}},
\end{equation}
where
\begin{eqnarray}\label{norms modes mu=1}
\nonumber
  N_L(\alpha) = \frac{\alpha!(\alpha+3)!}{72G}N_{log}^{-1}(\alpha)  \,\,\,\,\,\,\,\,\,\,\,\,\,\,\,\,\,\,\,\,\,  N^2_R(\bar{\alpha}) = \frac{2}{72G}\bar{\alpha}! (\bar{\alpha}+3)!
\ \ \ \ \ \ \ \ \ \ \ \ \ \ \ \ \ \
  \\
  N^2_{log}(\alpha) = \frac{\alpha! (\alpha +3)!}{216G}(3H(\alpha+3)-5) \,\,\,\,\,\,\,\, N^2_{GKP}(\alpha, \bar{\alpha}) = \frac{1}{72G} \alpha! \bar{\alpha}! (\bar{\alpha}-1)! (\alpha+3)!, \ \ \ \
\end{eqnarray}
where $H(\alpha) = \sum_{k=1}^\alpha \frac{1}{k}$ is the harmonic number.
The symplectic products involving $\hat \Psi_{log}^\alpha, \hat \Psi_{log}^{\alpha, \bar \alpha}$  can either be calculated directly from the wave functions, or by using (\ref{psi log}) and the results for $\mu > 1$.  In either case one finds
\begin{equation}\label{prods normalized mu = 1}
    (\hat{\Psi}_R^{\bar{\alpha}},\hat{\Psi}_R^{\bar{\beta}} ) = \delta^{\bar{\alpha} \bar{\beta}} \,\,\,\,  (\hat{\Psi}_{log}^\alpha,\hat{\Psi}_L^\beta) = -\delta^{\alpha \beta} \,\,\,\, (\hat{\Psi}_{log}^\alpha,\hat{\Psi}_{log}^\beta) = -\delta^{\alpha \beta} \,\,\,\, (\hat{\Psi}_{log}^{\alpha \bar{\alpha}}, \hat{\Psi}_{log}^{\beta \bar{\beta}}) = -\delta^{\alpha \beta} \delta^{\bar{\alpha} \bar{\beta}},
\end{equation}
with all other products vanishing.  In particular, $(\hat{\Psi}_{L}^{\alpha }, \hat{\Psi}_{L}^{\beta }) =0$. The normalized fields for $\mu=1$ were decorated with a hat (instead of with a tilde), to distinguish them from the limits of the tilded fields for $\mu > 1$, in particular we have $\lim_{\mu \to 1} \tilde{\Psi}_M^{\alpha \bar{\alpha}} = \hat{\Psi}_M^{\alpha \bar{\alpha}}$, but $\lim_{\mu \to 1} \tilde{\Psi}_L \neq  \hat{\Psi}_L$.

A few comments on (\ref{prods normalized mu = 1}) are in order. First, we emphasize that the products involving $\Psi_{log}$ are finite, so $\Psi_{log}$ is indeed a normalizable mode with respect to (\ref{omega TMG}). Second, we see that $\Psi_{log}$ has negative norm.  We therefore refer to $\Psi_{log}$ as a ghost\footnote{Here we abuse the terminology somewhat.  Due to the logarithmic behavior, $\Psi_{log}$ is not strictly-speaking a positive frequency mode.  However, it is useful to think of it as being effectively so since it is the limit of positive frequency modes for $\mu > 1$.}. This might be expected from the fact \cite{grumiller log} that $\Psi_{log}$ is known to carry negative energy, though the connection is not direct due to the complicated time-dependence.
Finally, despite the fact that $\Psi_L$ has vanishing norm, this mode cannot be discarded as pure gauge since it has non-zero symplectic product with $\Psi_{log}$.

\section{The linearized unitary quantum theory}
\label{newQ}

With the results from section \ref{prelim} in hand, we are ready to quantize linearized TMG; see also  \cite{canonical quantization of TMG buchbinder,canonical quantization TMG deser xiang} for other studies of quantum Topologically Massive Gravity.  We first analyze the case $\mu > 1$ and then study the log gravity limit $\mu  \to 1$.

The standard procedure for an operator quantization is to expand the general linear solution in some basis of modes.  The coefficients of these modes are quantum operators whose commutation relations are determined by the symplectic structure.  One then uses these operators to define a vacuum state, in the sense of a state with no particles, and thence an entire Fock space.  Note that for TMG the Hamiltonian is unbounded below for all $\mu$,  so that minimizing the energy does not lead to any preferred notion of vacuum state.

The above procedure is most familiar in the case where the modes diagonalize the symplectic structure, though the presence of ghosts brings certain subtleties.
For example, a mode expansion of the form $\Psi = b^\dag \psi + b \psi^*$ with $(\psi, \psi)= \pm1$ leads to $[b, b^\dag] = \pm 1$ \cite{Wald}. Recall that, at the level of the mode expansion, it is merely a matter of convention which coefficient is called $b$ and which is called $b^\dag$.  However it is useful to choose a convention which will lead to familiar expressions in the associated construction of a Fock space.  To this end, we rename $(b, b^\dag)$ as either $(a, a^\dag)$ or $(a^\dag, a)$ in such a way that the operator $a$ will annihilate the desired vacuum state $|0 \rangle$.  In order that $a^\dag |0 \rangle$ have positive norm, we must take $b = a, b^\dag = a^\dag$ for $(\psi, \psi)= 1$, but $b = a^\dag, b^\dag = a$ for $(\psi, \psi)=  -1$.  The resulting Fock space then defines a unitary quantization of the theory.  Note that this discussion applies regardless of whether $\psi$ is a positive frequency mode, a negative frequency mode, or of indeterminate frequency.  We shall use this rule when defining our mode expansion below.

On the other hand, for ghost modes like $\tilde \Psi_M$ (for $\mu > 1$), there is a different convention that might also have been considered natural. Following the convention for normal particles, one might take the coefficient of the positive frequency mode $\tilde \Psi_M$ to be $a^\dag_{M,NU}$ and that of the negative frequency mode $\tilde \Psi_{M}^*$ to be $a_{M,NU}$.  However, this would lead to  $[a_{M,NU}, a_{M,NU}^\dag] =  - 1$.    As a result, the Fock space defined over a vacuum that satisfies  $a_{M,NU} |0 \rangle = 0$ would contain negative norm states.  Indeed, it is clear that making such a choice for any of the ghosts leads to a non-unitary quantization  -- thus the subscript $NU$ on $a^\dag_{M,NU}$ above.  However we reserve the term {\it the non-unitary quantization} for the quantization scheme in which {\it all} the positive frequency modes are associated with creation operators (when at least one of them is a ghost). In the reminder of this section, we consider only unitary quantization schemes for log gravity and TMG, saving discussion of the non-unitary scheme for section \ref{NU}.

For $\mu =1$, the modes described in section \ref{prelim} do not diagonalize the symplectic structure (see \ref{prods normalized mu = 1}).  As a result, while one may consider the associated mode expansion

\begin{equation}\label{naive mode expansion psi log psi L}
    \Psi = \sum_{\alpha = 0} [ \hat a^\dag _{L \alpha} \hat{\Psi}_L^\alpha + a^\dag _{log \alpha} \hat{\Psi}_{log}^\alpha] +
    \sum_{\alpha = 0 }\sum_{\bar{\alpha} = 1} a_{M \alpha \bar{\alpha}} \hat{\Psi}^{\alpha \bar{\alpha}}_{log}  + \sum_{\bar{\alpha}=0} a^\dag_{R \bar{\alpha}} \hat{\Psi}^{\bar{\alpha}}_R +  h.c.,
\end{equation}
it is not immediately clear how to use  $\hat a_{L \alpha}$, $a_{log \alpha}$, $a_{M \alpha \bar{\alpha}}$ and their adjoints to define a useful (unitary) vacuum\footnote{In fact,
since $(\Psi^\alpha_L, \Psi^\alpha_{log} ) \neq 0$, it is not clear whether there is any advantage to refering to  coefficients of these modes as $a$ or $a^\dag$ in the mode expansion, but we make the choices above.}.  One needs to first diagonalize the symplectic structure and then apply the rule above.  We will do so below in a way that demonstrates the continuity of the unitary quantization scheme at $\mu =1$. In particular, though log gravity is often said to be non-unitary, we describe a unitary quantization below.  In (\ref{naive mode expansion psi log psi L}), the hat on $\hat a^\dag _{L \alpha}$ distinguishes this operator from another similar operator that will be greater use below.

\subsection{A family of vacuum states}
\label{section choice of vacuum state}

We begin with the case  $\mu > 1$, where it is natural to expand in a basis of modes with  well-defined conformal weights; i.e., in the basis  (\ref{list of modes and descendants gen mu}).  Following the convention described above for a unitary quantization scheme, we have
\begin{equation}\label{mode expansion generic mu well defined conformal weights}
    \Psi = \sum_{\alpha =0} [ a^\dag_{L \alpha} \tilde{\Psi}_L^\alpha(\mu) +  a_{M \alpha} \tilde{\Psi}_{M}^{\alpha}(\mu) ] +  \sum_{\alpha=0} \sum_{\bar{\alpha}=1}  a_{M \alpha \bar{\alpha}} \tilde{\Psi}_{M}^{\alpha \bar{\alpha}}(\mu)  + \sum_{\bar{\alpha}=0} a^\dag_{R \bar{\alpha}} \tilde{\Psi}_R^{\bar{\alpha}}(\mu)  +    h.c.
\end{equation}
Using this mode expansion, one may define a state $|0\rangle_\mu^{conf}$ annihilated by $a_{L \alpha}, a_{R \alpha}, a_{M \alpha \bar \alpha}$.  We refer to this state as the conformal vacuum, as it will turn out to be annihilated by all quantum charges $L_n, \bar L_n$ for $n \ge -1$.

However, due to the fact that (\ref{mode expansion generic mu well defined conformal weights}) degenerates as $\mu \to 1$, the conformal vacuum becomes singular in this limit.  Indeed, the two-point function defined by (\ref{mode expansion generic mu well defined conformal weights}) becomes a sum of divergent terms  due to the normalization factors (\ref{norms generic mu}) used to defined the tilded-fields (\ref{NormalizedGeneric}).  More definitively, one may note that since gauge transformations are degenerate directions of the symplectic structure, the operator
$\Omega(\Psi, \Psi_{log \alpha})$ is gauge invariant and satisfies

\begin{equation}
\label{0div}
{}_\mu \langle 0 | \Omega(\Psi, \Psi_{log}^\alpha) [\Omega(\Psi_{log}^\beta ,\Psi)]^\dag  | 0 \rangle_{\mu} = \frac{\alpha! P(4, \alpha) }{12G \mu(\mu-1)} \delta_{\alpha \beta}  ,
\end{equation}
In (\ref{0div}) we have defined  $\Psi^\alpha_{log}(\mu) = \frac{1}{\mu-1}(\Psi^\alpha_M - \Psi^\alpha_L) $ for $\mu \neq 1$, which of course gives just (\ref{psi log}) as $\mu \to 1$.

As a result, to find a construction of the TMG Hilbert space that is continuous at $\mu=1$ it will be useful to consider linear combinations that mix $\Psi^\alpha_L$ and $\Psi_M^\alpha$. For simplicity we superpose only modes with the same value of $\alpha$.  It is also useful to  keep the symplectic structure diagonal.  Since the matrix of symplectic products in the basis $\tilde{\Psi}_M^\alpha, \, \tilde{\Psi}_L^\beta$,  takes the form
\begin{equation}\label{matrix simp products psiM psiL = eta}
    (\tilde{\Psi}_A^\alpha, \tilde{\Psi}_B^\beta) = \left(
                                        \begin{array}{cc}
                                          -1 & 0 \\
                                          0 & 1 \\
                                        \end{array}
                                      \right),
\end{equation}
we consider modes $\chi_1^\alpha, \chi_2^\alpha,$ given by
\begin{equation}\label{chi 1 chi 2 generic mu}
    \left(
      \begin{array}{c}
        \chi^\alpha_1(\mu) \\
        \chi^\alpha_2(\mu) \\
      \end{array}
    \right) \equiv
      S_\alpha(\mu)\left(
                   \begin{array}{c}
                     \tilde{\Psi}_M^\alpha \\
                     \tilde{\Psi}_L^\alpha \\
                   \end{array}
                 \right),
\end{equation}
where
\begin{equation}\label{S alpha mu}
    S_\alpha(\mu) = \left(
                    \begin{array}{cc}
                      \cosh y & -\sinh y \\
                      -\sinh y  & \cosh y  \\
                    \end{array}
                  \right)
\end{equation}
is an $Sp(2)$ transformation  and $y$ carries the dependence on $\alpha$ and $\mu$. We could also include reflections of the form $diag(-1,1)$ in (\ref{S alpha mu}), but this does not add anything interesting.  A unitary quantization corresponds to taking the annihilation operators to be the coefficients of $\chi^\alpha_1(\mu)$ and $\tilde{\psi}_M^{\alpha \bar{\alpha}}(\mu)$, and the creation operators to be coefficients of $\chi^\alpha_2(\mu)$ and $\tilde{\Psi}_R^{\bar{\alpha}}(\mu)$ in the mode expansion.  Note that each function $y_\alpha(\mu)$ defines a vacuum $|0\rangle_\mu$ for each $\mu$. In order for $|0 \rangle_\mu$ to define the same Hilbert space as $|0 \rangle^{conf}_\mu$, we must have  \cite{BD}
\begin{equation}
\label{y -> 0}
\sum_\alpha \sinh^2 y < \infty,
\end{equation} so that in particular  $\mathop {\lim }\limits_{\alpha \to \infty } y_\alpha(\mu) = 0$.     We have already remarked that $|0 \rangle_\mu$ cannot be a state of minimum energy, since the energy is unbounded below.  However, we also warn the reader that, because the modes $\chi_1^\alpha, \chi_2^\alpha$ contain superpositions of positive and negative frequencies, the vacuum $|0\rangle_\mu$ will not even be an energy eigenstate.

We are now ready to study the limit $\mu \to 1$. To do so, we need a class of $y_\alpha(\mu)$ for which the modes (\ref{chi 1 chi 2 generic mu}) are continuous in $\mu$ and, in particular, define  non-degenerate linear combinations of $\Psi_L$ and $\Psi_{log}$ that diagonalize $\Omega$ at $\mu=1$.   Choosing $y(\alpha,\mu)$ itself to be continuous at $\mu =1$ does not achieve our goal.  This would simply give a linear combination of the modes $\Psi_L, \Psi_M$ and  their descendants, while these modes are known to degenerate for $\mu =1$.  However, a simple choice of $S_\alpha(\mu)$ that satisfies these requirements  is
\begin{equation}\label{ansatz for S}
    e^{2y_\alpha(\mu)} = 1 + \frac{8 \gamma(\alpha)}{\mu-1} \,\,\, , \,\,\,\,\, {\rm when} \,\,\,\, \frac{\gamma(\alpha+1)}{\gamma(\alpha)} \rightarrow 0.
\end{equation}
In fact, inserting (\ref{ansatz for S}) into (\ref{S alpha mu}) and expanding in powers of $(\mu-1)$, we obtain
\begin{eqnarray}\label{chi1 chi2 for mu=1 in terms of q}
\nonumber
    \chi_1^\alpha (\mu=1) &=& e^{q(\alpha)}\hat{\Psi}^\alpha_{log} - \sinh q(\alpha) \hat{\Psi}^\alpha_{L}, \\
    \chi_2^\alpha (\mu=1) &=& -e^{q(\alpha)}\hat{\Psi}^\alpha_{log} + \cosh q(\alpha) \hat{\Psi}^\alpha_{L},
\end{eqnarray}
\noindent where
\begin{equation}\label{definition of q(a)}
  e^{q(\alpha)} = \sqrt{\frac{\gamma(\alpha)}{3\gamma_1(\alpha)}}  \,\,\, , \,\,\,\,\,\,\,\,\,  \gamma_1(\alpha) = \frac{1}{2(-5+3H(\alpha+3))}.
\end{equation}
Note that $\gamma_1(\alpha) > 0$ for all $\alpha \geq 0$. Using (\ref{prods normalized mu = 1}), we can readily check that  $(\chi_i^\alpha (\mu=1), \chi_j^\beta (\mu=1)) = \delta_{ij} \delta^{\alpha \beta} (-1)^j$ as desired.

The remaining modes behave very simply as $\mu \to 1$.  It is manifest that $\Psi_R$ is continuous at $\mu =1$.  This leaves only $\tilde{\Psi}_M^{\alpha \bar{\alpha}}(\mu)$ for which, after a gauge transformation, the  $\mu \to 1$ limit turns out to be just  $\hat{\Psi}^{\alpha \bar{\alpha}}_{log}$.  This is most easily seen by writing
\begin{equation}\label{definition chi 3}
    \tilde{\Psi}_M^{\alpha \bar{\alpha}}(\mu) \approx \left[ \frac{\mu - 1}{ N_M(\alpha, \bar{\alpha}, \mu) } \right ] \frac{(\Psi_M^{\alpha \bar{\alpha}}(\mu) - \Psi_L^{\alpha \bar{\alpha}})}{\mu - 1} = \hat{\Psi}^{\alpha \bar{\alpha}}_{log} + {\cal O}(\mu-1),
\end{equation}
where $\approx$ means equality up to pure gauge modes (and we have used the fact that
$\Psi_L^{\alpha \bar{\alpha}}$ is pure gauge for all $\mu$).  Thus, as desired, $\chi_1^\alpha, \chi_2^\alpha, \Psi_R^\alpha, {\Psi}_M^{\alpha \bar{\alpha}}$ and their complex conjugates form a complete set of modes which is continuous  at $\mu =1$, at least up to gauge transformations. In our unitary quantization scheme, the mode expansion reads\footnote{It is maybe clearer to refer to the coefficient of the modes $\hat \Psi_{log}^{\alpha \bar \alpha}$ as $a_{\log}$ or $a_3$ for $\mu=1$ despite the fact that they are the $\mu \to 1$ limit of $a_{M \alpha \bar{\alpha}}$.}
\begin{eqnarray}\label{mode expansion for mu=1 as limit}
    \Psi &=& \sum_{\alpha =0} [ a_{1 \alpha} \chi^\alpha_1(\mu) + a^\dag_{2 \alpha} \chi^\alpha_2(\mu) ] +
    \sum_{\alpha=0}\sum_{\bar{\alpha}=1} a_{M \alpha \bar{\alpha}} \tilde{\Psi}_M^{\alpha \bar{\alpha}}(\mu) + \sum_{\bar{\alpha}} a^\dag_{R \bar{\alpha}} \tilde{\Psi}^{\bar{\alpha}}_R(\mu) + h.c. \\
    &=&  \sum_{\alpha =0} [ a_{1 \alpha} \chi^\alpha_1(1) + a^\dag_{2 \alpha} \chi^\alpha_2(1) ] +
    \sum_{\alpha=0}\sum_{\bar{\alpha}=1} a_{M \alpha \bar{\alpha}} \hat{\Psi}_{log}^{\alpha \bar{\alpha}} +  \sum_{\bar{\alpha}=0} a^\dag_{R \bar{\alpha}} \hat{\Psi}^{\bar{\alpha}}_R \,\,\, + h.c. \,\,\,  + {\cal O}(\mu-1),
 \ \ \ \ \ \ \  \ \   \label{LogModes}
\end{eqnarray}
where  we see from (\ref{naive mode expansion psi log psi L}) and   (\ref{chi1 chi2 for mu=1 in terms of q}) that for $\mu=1$ we have
\begin{equation}\label{a1 a2 in terms of anew aL}
  a_{1 \alpha} = \cosh q(\alpha) a^\dag_{log \alpha} + e^{q(\alpha)} a^\dag_{L \alpha} \,\,\,\,\,\, a_{2 \alpha} = \sinh q(\alpha) a_{log \alpha} + e^{q(\alpha)} a_{L \alpha}.
\end{equation}
Since the algebra defined by $a_{1\alpha}, a_{2\alpha}, a_{M \alpha \bar \alpha}, a_{R \bar \alpha} $ and their adjoints is the same for all $\mu$,  one may think of the operators as being $\mu$-independent;   all of the $\mu$ dependence is carried by the modes.

We now turn to the
vacuum state $|0\rangle_\mu$ annihilated by $a_{1\alpha}, a_{2\alpha}, a_{M \alpha \bar \alpha}, a_{R \bar \alpha} $.  Since the modes themselves are continuous (up to gauge transformations),  all gauge-invariant correlation functions at separated points are also continuous at $\mu=1$.  In this sense the vacuum $|0\rangle_\mu$ is itself  continuous at $\mu =1$.     Again, we remind the reader that such continuity does not hold for the conformal vacuum defined by (\ref{mode expansion generic mu well defined conformal weights}), though the latter is a highest-weight state.  As for $|0 \rangle_\mu$ when  $\mu > 1$, the $\mu \to 1$ limit is not a state of minimum energy and will not even be an energy eigenstate, much less a highest weight state.

\section{Charges and constraints}
\label{CandC}

As discussed above, our unitary quantization scheme defines a vacuum state (and thus an entire TMG Hilbert space) that is continuous at $\mu=1$.   There we take it to define a unitary quantization of log gravity.  While the right-moving charges $\bar L_n$ and most of the left-moving charges $L_n$ are also continuous, we will find in section \ref{section quantum charges for generic mu} below that $L_{\pm 1}$ are not.  Indeed, their action on $|0\rangle_\mu$ diverges as $\mu \to 1$.  In contrast, section \ref{NU} shows that all charges are continuous in the non-unitary quantization.

\subsection{Virasoro charges}
\label{section quantum charges for generic mu}

At the classical level, one may build conserved charges directly from the symplectic structure.  In general, given a vector field $\xi$, the infinitesimal difference between the associated charges of solutions $\bar{\phi}$ and $\bar{\phi} + \delta \phi$ is
\begin{equation}\label{delta Q definition}
    \delta_1 Q_\xi = \Omega(\delta_1 \phi , ({\cal L}_\xi \bar{\phi})^*; \bar{\phi} ).
\end{equation}
In the linearized theory, one may hold the background $\bar \phi$ fixed once and for all.
If ${\cal L}_\xi \bar{\phi} = 0$, (\ref{delta Q definition}) can then be integrated to yield a quadratic expression in $\delta \phi$:
\begin{equation}\label{Q integrated general}
    Q_\xi(\delta_1 \phi) = \frac{1}{2} \Omega(\delta_1 \phi , ({\cal L}_\xi \delta_1 \phi)^*; \bar{\phi} ).
\end{equation}
On the other hand, if $\xi = \xi_{asympt}$ is not an exact symmetry of $\bar \phi$ but only an asymptotic symmetry, the conserved charge in the linearized theory is just
\begin{equation}\label{Q asympt general}
    Q_\xi(\delta_1 \phi) = \Omega(\delta_1 \phi , ({\cal L}_{\xi_{asympt}} \bar{\phi})^*; \bar{\phi} ).
\end{equation}

At the quantum level, we promote  (\ref{Q integrated general}) and (\ref{Q asympt general}) to operators by expressing them  in terms of the coefficients of our mode expansion.   Of course, one must choose an appropriate ordering of operators.  In making this choice, one wishes to preserve the classical symmetries.  This means that the charges should be conserved, should satisfy the Virasoro algebra, and should be invariant under the discrete symmetry $\Psi \to -\Psi$.  In terms of the mode expansion (\ref{mode expansion generic mu well defined conformal weights}) associated with the conformal vacuum for $\mu > 1$, it is sufficient to simply normal-order the classical expression (\ref{Q integrated general}) and to replace $\delta_1 \phi$ by the linearized quantum field $\Psi$ in (\ref{Q asympt general}); i.e., we have
\begin{eqnarray}\label{bar L0 generic mu}
    \bar{L}_0 &=& \sum_{\bar{\alpha}=0} (2+\bar{\alpha}) a^\dag_{R \bar{\alpha}} a_{R \bar{\alpha}} - \sum_{\alpha=0} \sum_{\bar{\alpha}=0}(-\frac{1}{2} + \frac{\mu}{2} + \bar{\alpha}) a^\dag_{M \alpha \bar{\alpha}} a_{M \alpha \bar{\alpha}}, \\
    \label{bar L1 generic mu}
    \bar{L}_{-1} &=& \sum_{\bar{\alpha}=0} \sqrt{(\bar{\alpha}+1)(\bar{\alpha}+4)} a^\dag_{R (\bar{\alpha}+1)} a_{R \bar{\alpha}} - \sum_{\alpha=0} \sum_{\bar{\alpha}=0}\sqrt{(\bar{\alpha}+1)(-1+\mu+\bar{\alpha})} a^\dag_{M \alpha \bar{\alpha}} a_{M \alpha (\bar{\alpha}+1)}, \ \ \ \ \ \ \ \ \  \\
\label{barL -alpha alpha1}
    \bar{L}_{\bar{\alpha}} &=& \frac{3}{(\bar{\alpha}-2)!} N_R(\bar{\alpha}-2,\mu)a_{R (\bar{\alpha}-2)}, \ {\rm for } \ \bar \alpha \ge 2
\end{eqnarray}
  \begin{eqnarray}\label{L0 generic mu}
    {L}_0 &=& \sum_{{\alpha}=0} (2+{\alpha}) a^\dag_{L {\alpha}} a_{L {\alpha}} - \sum_{\alpha=0} \sum_{\bar{\alpha}=0}(\frac{3}{2} + \frac{\mu}{2} + {\alpha}) a^\dag_{M \alpha \bar{\alpha}} a_{M \alpha \bar{\alpha}}, \\
    \label{L1 generic mu}
    {L}_{-1} &=& \sum_{{\alpha}=0} \sqrt{({\alpha}+1)({\alpha}+4)} a^\dag_{L ({\alpha}+1)} a_{L {\alpha}} - \sum_{\alpha=0} \sum_{\bar{\alpha}=0}\sqrt{({\alpha}+1)(3+\mu+{\alpha})} a^\dag_{M \alpha \bar{\alpha}} a_{M (\alpha+1) \bar{\alpha}}, \ \ \ \ \ \ \ \ \ \\
    \label{L -alpha alpha1}
    {L}_{{\alpha}} &=& \frac{3}{({\alpha}-2)!} N_L({\alpha}-2,\mu)a_{L ({\alpha}-2)}, \ {\rm for } \ \alpha \ge 2,
\end{eqnarray}
with the other charges determined by $L_{-\alpha} = L_\alpha^\dagger$.

In particular, the algebra forbids us from adding further c-number constants.   The success of normal ordering (without additional c-number terms) can be shown to follow from the fact that each creation or annihilation operator appearing in (\ref{bar L0 generic mu}), (\ref{bar L1 generic mu}) has a well-defined conformal weight greater than $1/2$.  We note for future reference that although the quadratic operators are represented as sums over an infinite number of modes, these sums converge converge in the Hilbert space norm when the operators act on a Fock space state an the appropriate domain.  We take this domain to include the dense linear subspace $\Phi^{conf}$ defined by the conformal vacuum and all states obtained from it by adding a finite number of particles. The choice of vacuum thus defines a regulator that gives meaning to any potentially ill-defined expressions arising from these infinite sums. Expressions (\ref{bar L0 generic mu}, \ref{bar L1 generic mu}, \ref{L0 generic mu}, \ref{L1 generic mu}) and their adjoints satisfy the $SL(2,R)$ algebra in this sense.  The larger Virasoro algebra also holds, with the caveat that since in this approximation the Virasoro charges $L_\alpha, \bar L_{\bar \alpha}$ for  $|\alpha| \geq 1$ are linear in the fields, the commutator of two such charges gives us only the central charge term (\ref{central charges generic mu}) in the Virasoro algebra.

However, as noted earlier, the conformal vacuum $|0 \rangle^{conf}$ (and thus the entire space $\Phi^{conf}$) becomes singular at $\mu =1$.  To study the $\mu \to 1$ limit, we should thus use a different dense linear space $\Phi$ associated with the vacuum $|0\rangle_\mu$ defined by some $y_\alpha(\mu)$ as in section \ref{newQ} and including the states obtained from it by acting with a finite number of creation operators from the mode expansion (\ref{mode expansion for mu=1 as limit}).  Using
\begin{equation}
a_{L\alpha} = -\sinh y_\alpha \ a_{1\alpha}^\dag + \cosh y_\alpha \  a_{2\alpha},  \ \ \
a_{M\alpha} = \cosh y_\alpha \ a_{1\alpha} - \sinh y_\alpha a_{2\alpha}^\dag,
\end{equation}
one may verify that for $\mu > 1$ the action of (\ref{bar L0 generic mu}-\ref{barL -alpha alpha1}) on $|0\rangle_\mu$ is well-defined so long as $y_\alpha(\mu) \to 0$ fast enough as $\alpha \to \infty$; e.g., if (\ref{ansatz for S}) holds. In fact, they are well-defined on all of $\Phi$.  The results of such calculations are summarized by writing the charges in terms of the mode expansion (\ref{mode expansion for mu=1 as limit}) as follows.  The right-moving $SL(2,R)$ charges are given by
\begin{eqnarray}
\nonumber
  \bar{L}_0 &=& \sum_{\bar{\alpha}=0} (2+\bar{\alpha}) a^\dag_{R \bar{\alpha}} a_{R \bar{\alpha}} - \sum_{\alpha=0} \sum_{\bar{\alpha}=1}(\frac{1}{2} - \frac{\mu}{2} + \bar{\alpha}) a^\dag_{M \alpha \bar{\alpha}} a_{M \alpha \bar{\alpha}} - \frac{1}{2}(\mu-1) \sum_{\alpha=0} \bigg \{ \cosh^2 y \ a_{1\alpha}^\dag a_{1\alpha}   \\
  \nonumber
   &+& \sinh^2 y \ a_{1\alpha}^\dag a_{1\alpha} - \cosh y \sinh y ( a^\dag_{1\alpha} a^\dag_{2\alpha} + a_{1\alpha} a_{2\alpha}     )  \bigg \} + \bar{C}_0 \\
\nonumber
    \bar{L}_{-1} &=& -(\mu-1)^{1/2} \sum_{\alpha=0} a_{M \alpha 1}( \cosh y a^\dag_{1 \alpha} -  \sinh y a_{2 \alpha}   )
    - \sum_{\alpha=0} \sum_{\bar{\alpha}=1}\sqrt{({\alpha}+1)(-1+\mu+{\alpha})} a^\dag_{M \alpha \bar{\alpha}} a_{M (\alpha+1) \bar{\alpha}}  \\
   &+&\sum_{\bar{\alpha}=0} \sqrt{(\bar{\alpha}+1)(\bar{\alpha}+4)} a^\dag_{R (\bar{\alpha}+1)} a_{R \bar{\alpha}}
  \label{right12}
\end{eqnarray}
and their adjoints, where
\begin{equation}\label{bar C0}
    \bar{C}_0 = - \frac{(\mu-1)}{2} \sum_{\alpha=0} \sinh^2 y.
\end{equation}
 In contrast, since our $Sp(2)$ transformation does not act on the modes $\Psi_{R\bar \alpha}$,
the higher right-moving Virasoro charges are unchanged.  Finally, the left-moving charges become
\begin{eqnarray}\label{l0 generic mu}
\nonumber
  L_0(\mu) &=& \sum_{\alpha =0} \bigg \{ a^\dag_{1 \alpha} a_{1 \alpha} ( h_L^\alpha \sinh^2 y - h_M^\alpha \cosh^2 y ) + a^\dag_{2 \alpha} a_{2 \alpha} ( h_L^\alpha \cosh^2 y - h_M^\alpha \sinh^2 y )  \\
  \nonumber
   &+& (a_{1\alpha} a_{2 \alpha} +  a^\dag_{1\alpha} a^\dag_{2 \alpha}  ) \sinh y \cosh y (h_M^\alpha - h_L^\alpha) \bigg \} - \sum_{\alpha=0} \sum_{\bar{\alpha}=1} (\frac{3}{2}+ \frac{\mu}{2} +\alpha) a^\dag_{M \alpha \bar{\alpha}} a_{M \alpha \bar{\alpha}} + C_0 \ \ \ \ \  \ \ \ \ \\
   \nonumber
       L_{-1}(\mu) &=& \frac{1}{2} \sum_{\alpha=0} \bigg \{ A_{--}(\mu,\alpha) a^\dag_{1\alpha} a_{1 (\alpha+1)} + A_{++}(\mu,\alpha) a^\dag_{2(\alpha+1)} a_{2 \alpha} + A_{+-}(\mu,\alpha) a_{2\alpha} a_{1 (\alpha+1)}  \\
       \nonumber &+& A_{-+}(\mu,\alpha) a^\dag_{1\alpha} a^\dag_{2 (\alpha+1)} \bigg \}
    - \sum_{\alpha=0} \sum_{\bar{\alpha}=1} \sqrt{ (\alpha+1)(\alpha + 3 + \mu) } a^\dag_{M \alpha \bar{\alpha}} a_{M (\alpha+1) \bar{\alpha}} \\
    \label{right12}
    L_{\alpha} &=& \frac{3}{(\alpha-2)!} N_L(\alpha-2, \mu) \bigg \{ \cosh[y_{\alpha-2}(\mu)] a_{2 (\alpha-2)} - \sinh[y_{\alpha-2}(\mu)] a^\dag_{1 (\alpha-2)}  \bigg \}.
\end{eqnarray}
and their adjoints, where
$h_L^\alpha = 2+\alpha \,\,\,   h_M^\alpha = \frac{3}{2} + \frac{\mu}{2} + \alpha$, $C_0  = \bar C_0$,
and
\begin{equation}\label{Aij constants for generic mu}
  A_{ij}(\mu,\alpha)  = 2(-1)^{i j} (\alpha +1)^{1/2} \left( (2h_L + \alpha)^{1/2}  f_i(y_\alpha) f_j (y_{\alpha+1}) - (2h_M + \alpha)^{1/2} f_{-i}( y_\alpha) f_{-j} (y_{\alpha+1}) \right) \ \ \
  \end{equation}
for $i,j \in \{-,+\}, \  f_- (y)  = \sinh(y), \ f_+(y) = \cosh(y).$
Finiteness for $\mu >1$ follows from the fact that (\ref{bar C0}) converges absolutely and, since (\ref{ansatz for S}) requires $y_\alpha$ to vanish faster than any exponential at large $\alpha$, from the fact that the coefficient of each term involving only creation operators includes $\sinh y_\alpha$ .  Note that due to the terms of the form $a^\dag_{1 \alpha} a^\dag_{2 \alpha}$ in $L_0, \bar L_0$, our vacuum $|0\rangle_\mu$ is not an eigenstate of either generator, nor of the energy $L_0 + \bar L_0$, though it is a state of zero angular momentum: $(L_0 - \bar L_0 ) |0 \rangle_\mu =0$.
This structure may seem somewhat artificial for generic $\mu$, but is an intrinsic property of log-gravity since for $\mu =1$ it is impossible to diagonalize the action of  $L_0$ on the space of linearized solutions. Indeed, this  property  was noted in \cite{grumiller log} and used there to suggest a connection with logarithmic CFTs  \cite{LCFT log operators in CFT gurarie}, see \cite{LCFT bits and pieces flohr} and \cite{LCFT algebraic approach gaberdiel} for reviews.

It is now straightforward to take the limit $\mu \to 1$. Using (\ref{ansatz for S}), one finds that all coefficients in (\ref{right12}), (\ref{right12}) are finite.  The non-trivial results are
\begin{eqnarray}
\nonumber
  \bar{L}_0(\mu=1) &=& \sum_{\bar{\alpha}=0} (2+\bar{\alpha}) a^\dag_{R \bar{\alpha}} a_{R \bar{\alpha}} - \sum_{\alpha=0} \sum_{\bar{\alpha}=1} \bar{\alpha} a^\dag_{M \alpha \bar{\alpha}} a_{M \alpha \bar{\alpha}} - \sum_{\alpha=0} \gamma(\alpha) \bigg \{ a_{1\alpha}^\dag a_{1\alpha}   \\
  \nonumber &+&  a_{1\alpha}^\dag a_{1\alpha} -  a^\dag_{1\alpha} a^\dag_{2\alpha} - a_{1\alpha} a_{2\alpha}  \bigg \} + \bar{C}_0 \\
\nonumber     \bar{L}_{-1}(\mu=1)   &=& - \sum_{\alpha=0} (2\gamma(\alpha))^{1/2} a_{M \alpha 1}( a^\dag_{1 \alpha} -  a_{2 \alpha} )
    - \sum_{\alpha=0} \sum_{\bar{\alpha}=1}\sqrt{\alpha({\alpha}+1)} a^\dag_{M \alpha \bar{\alpha}} a_{M (\alpha+1) \bar{\alpha}} \\
     &+&\sum_{\bar{\alpha}=0} \sqrt{(\bar{\alpha}+1)(\bar{\alpha}+4)} a^\dag_{R (\bar{\alpha}+1)} a_{R \bar{\alpha}}
\end{eqnarray}
\begin{eqnarray}
\nonumber
    L_0(\mu=1) &=& \sum_{\alpha=0} \bigg \{ - ( 2 + \alpha + \gamma(\alpha)) a^\dag_{1 \alpha} a_{1 \alpha}  + ( 2 + \alpha - \gamma(\alpha)) a^\dag_{2 \alpha} a_{2 \alpha}
   \\ \nonumber &+& \gamma(\alpha) (a_{1\alpha} a_{2 \alpha} +  a^\dag_{1\alpha} a^\dag_{2 \alpha}  ) \bigg \}
   - \sum_{\alpha=0} \sum_{\bar{\alpha} =1} (2+\alpha) a^\dag_{M \alpha \bar{\alpha}} a_{M \alpha \bar{\alpha}} + C_0 \,\,\,\, , \\
   \nonumber
    L_{-1}(\mu=1) &=& \frac{1}{2} \sum_{\alpha=0} \bigg \{ A_{--}(\alpha) a^\dag_{1\alpha} a_{1 (\alpha+1)} + A_{++}(\alpha) a^\dag_{2(\alpha+1)} a_{2 \alpha} + A_{+-}(\alpha) a_{2\alpha} a_{1 (\alpha+1)} \\ \nonumber &+& A_{-+}(\alpha) a^\dag_{1\alpha} a^\dag_{2 (\alpha+1)} \bigg \}
    - \sum_{\alpha=0} \sum_{\bar{\alpha}=1} \sqrt{ (\alpha+1)(\alpha + 3 ) } a^\dag_{M \alpha \bar{\alpha}} a_{M (\alpha+1) \bar{\alpha}}, \\
    L_{\alpha} &=& \left[ \frac{\alpha(\alpha^2-1)\gamma(\alpha-2)}{4G} \right]^{1/2} ( a_{2 \alpha} - a^\dag_{1 \alpha} )  \ \ \ {\rm for} \ \alpha \ge 2.
\end{eqnarray}
\noindent where
\begin{equation}\label{bar C0}
    \bar{C}_0 =  C_0 = - \sum_{\alpha=0} \gamma(\alpha)
\end{equation}
and
\begin{equation}\label{Aij constants mu=1}
\nonumber
  A_{ij}(\alpha)  =  \left[ \frac{(\alpha+1) (\alpha+4)}{\gamma(\alpha+1)\gamma(\alpha)} \right]^{1/2} \left[  (-1)^j \gamma(\alpha+1) + (-1)^i \gamma(\alpha)   - 2 \frac{ (-1)^{ij}   \gamma(\alpha+1)\gamma(\alpha)}{\alpha + 4} \right].   \ \ \
  \end{equation}

For most of the above charges, acting on $|0 \rangle_\mu$ continues to give a normalizable state in this limit.  However, the norm of $L_{\pm 1}|0 \rangle$ diverges as $\mu \to 1$ as can be seen from the fact that the 2nd term in the coefficient $A_{-+}$, of the $a^\dag_{1 \alpha} a^\dag_{2 (\alpha+1)}$ terms in $L_{-1}$ now grows with $\alpha$ when $\gamma(\alpha)$ vanishes rapidly as $\alpha \to \infty$.  The same is true for the analogous coefficient $A_{+-}$  in $L_1$.  As a result, the $\mu = 1$ Hilbert space that defines our unitary quantization of log gravity appears to carry a representation of only the right-moving Virasoro algebra.  Taking $\gamma \to 0$ more slowly is not helpful, as one can show that $A_{-+} \to 0$ implies that $\gamma$ approaches a non-zero constant at large $\alpha$.

Note that, at the linearized level, the action of charges on the field operators is just the same as in the classical theory and so remains well-defined as $\mu \to 1$.  In this sense, the {\it theory} retains the left-moving Virasoro symmetry, though the symmetry is broken at the level of the Hilbert space. Symmetries of this sort are typically referred to as ``spontaneously broken,'' though as discussed in section \ref{disc} the fact that our case features spontaneous breaking of an {\em asymptotic} symmetry makes it somewhat different from more familiar cases of spontaneous symmetry breaking.

\subsection{The non-unitary quantization}
\label{NU}

Although our main focus is on unitary quantizations of TMG and log gravity, we now briefly discuss the situation for the non-unitary quantization.  This treatment largely coincides with that of \cite{Myung:2008dm} and may be considered a review. As noted in section \ref{newQ}, here one takes creation operators to be the coefficients of positive frequency modes and annihilation operators to be the coefficients of negative frequency modes in any expansion where each mode has a well-defined sign of the frequency.  One then defines a vacuum state $|0 \rangle^{NU}_\mu$ which is annihilated by the annihilation operators and uses the creation operators to build a Fock space which, in the presence of ghosts, will contain negative-norm states.   The details of the mode expansion do not affect the definition of $|0 \rangle^{NU}_\mu$, since any two allowed mode expansions are related by a transformation that maps creation operators to sums of creation operators and similarly for annihilation operators.  For $\mu =1$ one may safely classify the mode $\Psi_{log}$ as a positive-frequency mode since it is the limit of positive-frequency modes for $\mu > 1$.

The above invariance under changes of the mode expansion means that, while for $\mu > 0$ it is most natural to use a basis of modes with well-defined conformal weights given by (\ref{list of modes and descendants gen mu}) and their conjugates, and while this expansion degenerates at $\mu =1$, the corresponding vacuum state $|0 \rangle^{NU}_\mu$ remains continuous at $\mu=1$.  There it coincides with the non-unitary vacuum defined by (\ref{normalized modes mu = 1}) and their conjugates.  This continuity, combined with the well-defined conformal weights of (\ref{list of modes and descendants gen mu}) for $\mu > 1$, makes for a simple analysis.  The right- and left-moving charges take a form that is essentially that of (\ref{bar L0 generic mu} -  \ref{L -alpha alpha1}) with appropriate re-definitions of creation and annihilation operators.  In particular, it now suffices to take the quadratic operators to be normal-ordered with respect to the non-unitary creation/annihilation operators without adding any additional c-number terms\footnote{Note that this is not equivalent to simply rewriting (\ref{bar L0 generic mu} -  \ref{L -alpha alpha1}) in terms of the non-unitary creation and annihilation operators.  For example, the two definitions of $L_0$ differ by an infinite ordering constant.  The point here is that each Hilbert space defines a different notion in which the relevant mode sums should converge, and that these notions are not equivalent.}.  As a result, one finds $L_i^\dagger  | 0 \rangle_\mu^{NU} =0$, $\bar L_i^\dagger  | 0 \rangle_\mu^{NU} =0$ for all quadratic charges ($i = -1,0,1$) for all $\mu$.   The higher charges with $i \ge 2$ also annihilate the vacuum, and the charges with $i \le -2$ yield one-particle states with coefficients proportional to $N_L, N_R.$  No problems arise in the limit $\mu \to 1$.

It is thus straightforward to truncate the non-unitary $\mu=1$ theory using the left-moving charges.    In particular, one may impose the constraints

\begin{equation}
L_i |\psi \rangle =0
\end{equation}
for\footnote{It appears that similar constraints for $i \le -2$ have no solutions in the non-unitary Hilbert space.  This contrasts with the situation in the unitary Hilbert space, where the fact that $c_L =0$ means that when $L_i |\psi \rangle =0$  one also necessarily has $L_i^\dag |\psi \rangle =0$. See section \ref{disc} for further comments. } $i \ge -1$.    A natural space of solutions is given by the vacuum
$| 0 \rangle_\mu^{NU}$ and all $n$-particle states built by acting with all polynomials in creation and annihilation operators that commute with $L_i$.  Here it is useful to note that
due to
(\ref{Lminusa bar g prop Psi L}) and (\ref{prods normalized mu = 1}), the higher Virasoro charges (\ref{Q asympt general}) for $\mu =1$ take the form
\begin{eqnarray}\label{L -alpha = psi new}
    L_{\alpha} &=& - \sqrt{\frac{3\alpha(\alpha^2-1)}{4G}  \gamma_1(\alpha-2) } \ a_{log (\alpha-2)}, \ \ \ {\rm and}\\
         L_{-\alpha} &=& - \sqrt{\frac{3\alpha(\alpha^2-1)}{4G}  \gamma_1(\alpha-2) } \ a_{log (\alpha-2)}^\dag,
\end{eqnarray}
and that
inverting the symplectic structure (\ref{prods normalized mu = 1}) yields the commutators
\begin{equation}\label{commutators naive mode expansion psi new psi L}
    [a_{log \alpha}, a^\dag_{log \beta} ] = 0 \,\,\,\,\,\,\, [\hat a_{L \alpha}, \hat a^\dag_{L \beta} ] = + \delta_{\alpha \beta} \,\,\,\,\,\,\, [\hat a_{L \alpha}, a^\dag_{log \beta} ] = -\delta_{\alpha \beta} \,\,\,\,\, [a_{M \alpha \bar{\alpha}}^{NU}, a^{NU \dag}_{M \beta \bar{\beta}} ] =  - \delta_{\alpha \beta} \delta_{\bar{\alpha} \bar{\beta}},
  \end{equation}
  and of course $[a_{R\bar \alpha}, a^\dag_{R \bar \beta} ] = \delta_{\alpha \beta},$ where we have defined $a^{NU}_{M \alpha \bar \alpha} = a_{M \alpha \bar \alpha}^\dag.$
The desired polynomials are thus precisely those built from $a_{log \alpha}^\dag$ and $a_{R \alpha}^\dag$.  Due to the commutation relations (\ref{commutators naive mode expansion psi new psi L}), this leaves a positive semi-definite space of states.  Taking a quotient by the zero-norm states leaves a positive definite Fock space defined by acting on the vacuum with only the right-moving creation operators $a_{R \alpha}^\dag$.  This is just what one would expect from the classical theory of chiral gravity.  In particular, despite the non-unitary treatment of log gravity, this approach provides a unitary theory of chiral gravity as desired.

\section{Discussion}
\label{disc}

Our work above has studied the quantum theory of linearized anti-de Sitter topologically massive gravity for various values of the coupling $\mu \ell$.  Such theories generally contain ghosts.  As a result, while the classical theory is well-defined, the Hamiltonian is not bounded below.  Similarly, one may construct a well-defined unitary quantum theory (with positive probabilities) though the Hamiltonian is again unbounded below. At the classical level, the theory is continuous in $\mu \ell$ and one obtains the so-called log gravity theory by taking the limit as $\mu \ell$ approaches the chiral point ($\mu \ell \to 1$).   In the same way, both the quantum Hilbert space and local correlators at separated points are continuous at the chiral point, so that we have a unitary quantization of log gravity.

The above classical theories have both right- and left-Virasoro algebras of conserved charges which are again continuous at $\mu \ell =1$. Because we work in the linearized theory, the action of these charges on quantum fields is trivially the same as the action on classical fields and is again continuous at $\mu \ell=1$.  However, the action of two charges ($L_{\pm 1})$ on our vacuum state is not continuous and in fact diverges at $\mu \ell =1$.   As a result, only the right-moving Virasoro algebra is represented on the Hilbert space of our unitary quantization of log gravity.  The left-moving algebra may be said to be spontaneously broken.

Indeed, from the bulk point of view the phenomenon has much in common with more familiar cases of spontaneous symmetry breaking.  The divergence of $L_{\pm 1}$ on our unitary vacuum $|0 \rangle_\mu$ is an infra-red effect associated with the logarithmic behavior at the AdS boundary. This can be seen by replacing the vector fields $\xi_{\pm 1}$ defining $L_{\pm 1}$ with vector fields of compact support.  Because the associated charges generate gauge transformations, they vanish identically.  However, this also leads to an important difference: because the excitations generated by such truncated symmetries are pure gauge, one would not expect to find Goldstone bosons in the usual sense.   The point is that we find spontaneous breaking of an asymptotic symmetry, as opposed to a more conventional global symmetry.

We have worked at the level of the linearized theory, where the Virasoro algebra simplifies greatly.  Aside from the SL(2,R) algebra generated by $L_{\pm 1}, L_0$, it becomes essenitally a $U(1)$ current algebra.  At the non-linear level,  $L_{\pm 1}$ should appear in commutators of other left-moving charges.  Thus the divergence of $L_{\pm 1}$ implies that other left-moving charges must also diverge.  It is natural to expect that the left-moving Virasoro algebra is broken to just $L_0$.

Because the action of the left-moving symmetries became ill-defined at the chiral point, we could not define chiral gravity as the $L_n =0$ truncation of log gravity using our unitary quantization.  In contrast, no such difficulties arose in the non-unitary quantization where one obtained the expected chiral theory, which turns out to be unitary.

Since logarithmic conformal field theories are typically said to be non-unitary, the reader may wonder if our symmetry breaking in the unitary theory follows directly by an algebraic argument from the logarithmic structure of the primary fields.  The answer is not clear to us.  In particular, the usual argument for non-unitarity assumes that the vacuum is an eigenstate of $L_0$, a statement that is manifestly false in our unitary quantization.  Indeed, the usual argument for non-unitarity of logarithmic theories involves only $L_0$ (and not $L_{\pm 1}$), while we find this operator to be well-defined at $\mu \ell =1$.

From the perspective of chiral gravity, it would be very interesting to understand whether our breaking of the the left-moving symmetries in the unitary theory indicates a fundamental issue for quantum chiral gravity  or is merely an artifact of our construction.  There are in principle several logical possibilities, which we enumerate below.

The first possibility is that the unitary quantization of TMG for $\mu \ell > 1$
admits some more subtle $\mu \ell \to 1$ limit which defines a better behaved vacuum for log gravity.  Recall, for example, that for simplicity we considered only linear
transformations on the basis of mode functions that were diagonal in $\alpha$.    Perhaps mixing modes with different values of $\alpha$ would lead to better behavior for $L_{\pm 1}$?  For both the diagonal and the non-diagonal cases, it would be useful to understand better the behavior of the associated vacuum states in the far ultra-violet.  While we have noted that correlators in our vacuum $|0 \rangle_\mu$ are continuous at $\mu \ell =1$ when their arguments are separated, we have not studied the coincidence limits in detail. The fact that the our modes are well-behaved at short distances leads one to expect that the vacuum continues to have good short distance properties at $\mu=1$, but it would be useful to verify that composite operators can be renormalized in a useful way.

Another logical possibility is that there is some clever way to implement the constraints $L_n =0$ on our $\mu \ell =1$ Hilbert space despite the fact that some of the generators diverge. At the linearized level, it is straightforward to solve the $|n| \ge 2$ constraints using the fact (\ref{L -alpha = psi new}) that $L_{\alpha} \propto  a_{log (\alpha-2)}$.  Since at the linearized level we have $[L_n, L^\dagger_m] =0$ for $|n |, |m| \ge 2$, all of these operators may be simultaneously diagonalized in the unitary theory.  While the solutions to these constraints are not normalizable, they are easily controlled using the techniques of group averaging (see e.g. \cite{ALMMT,where}).  In fact, any solution of the constraints for $ n \ge 2$ necessarily also solves the constraints for $n \le -2$. Furthermore, at the classical level setting $a_{log  \alpha} =0 = a_{log \alpha}^\dag$ for $|n| \ge 2$ truncates the infinite sum that led to difficulties with $L_{\pm1}$.  It is therefore possible that there is a useful sense in which the remaining constraints $L_{\pm 1}=0, L_0 =0$ can be imposed on states solving the higher-order Virasoro constraints.  The problem, of course, is that to make use of the fact that $a_{log \alpha}, a_{log \alpha}^\dag$ annihilate the state, we must commute these operators to the right in the expressions for $L_{\pm 1}, L_0$.  Unfortunately, it is not clear to us how the infinite sums generated by this procedure can be controlled in a useful way.

A third logical possibility is that unitary theories of chiral gravity are simply not related to a unitary quantization of log gravity or of TMG for $\mu \ell > 1$.   For example, it may be that chiral gravity is best defined by truncating the non-unitary quantization of log gravity, or by using Brown-Henneaux boundary conditions to define the theory directly (without using log gravity as an intermediate step).  While such approaches give up any hope of connecting chiral gravity to a theory of quantum TMG with $\mu \ell \neq 1$ having positive probabilities, this might be justified by arguing that the presence of ghosts in log gravity or for $\mu \ell \neq 1$ suggests that chiral gravity is the only physically sensible theory resulting from TMG with asymptotically AdS boundary conditions.

The final logical possibility is that our breaking of the left-moving symmetry does in fact signal a fundamental issue for quantum chiral gravity.  Though we do not see a direct connection at this stage, it would be particularly interesting to relate this result to the arguments of \cite{gaberdiel no extremal cft} suggesting that extremal CFTs do not exist.

 \subsection*{Acknowledgements}
 The authors thank David Berenstein for useful discussions. During this work, TA was partly supported by a Fulbright-CONICYT fellowship. This work was also supported in part by the National Science Foundation under grants PHY05-55669 and PHY08-55415, and by funds from the University of California.


\begin{thebibliography}{10}

\bibitem{chiral 1 strominger}  W.~Li, W.~Song and A.~Strominger,
  ``Chiral Gravity in Three Dimensions,''
  JHEP {\bf 0804}, 082 (2008)
  [arXiv:0801.4566 [hep-th]].

\bibitem{AdS/CFT maldacena}  J.~M.~Maldacena,
  ``The large N limit of superconformal field theories and supergravity,''
  Adv.\ Theor.\ Math.\ Phys.\  {\bf 2}, 231 (1998)
  [Int.\ J.\ Theor.\ Phys.\  {\bf 38}, 1113 (1999)]
  [arXiv:hep-th/9711200].

\bibitem{AdS/CFT gubser klebanov polykov}  S.~S.~Gubser, I.~R.~Klebanov and A.~M.~Polyakov,
  ``Gauge theory correlators from non-critical string theory,''
  Phys.\ Lett.\  B {\bf 428}, 105 (1998)
  [arXiv:hep-th/9802109].


\bibitem{AdS/CFT witten}   E.~Witten,
  ``Anti-de Sitter space and holography,''
  Adv.\ Theor.\ Math.\ Phys.\  {\bf 2}, 253 (1998)
  [arXiv:hep-th/9802150].


\bibitem{chiral & log gravity and extremal cft strominger}  A.~Maloney, W.~Song and A.~Strominger,
  ``Chiral Gravity, Log Gravity and Extremal CFT,''
  arXiv:0903.4573 [hep-th].


\bibitem{witten 3d gravity solvable} E.~Witten,
  ``(2+1)-Dimensional Gravity as an Exactly Soluble System,''
  Nucl.\ Phys.\  B {\bf 311}, 46 (1988).

\bibitem{Witten partition function} A.~Maloney and E.~Witten,
  ``Quantum Gravity Partition Functions in Three Dimensions,''
  arXiv:0712.0155 [hep-th].

\bibitem{witten 3d gravity reconsidered} E.~Witten,
  ``Three-Dimensional Gravity Revisited,''
  arXiv:0706.3359 [hep-th].

\bibitem{gaberdiel no extremal cft} M.~R.~Gaberdiel, S.~Gukov, C.~A.~Keller, G.~W.~Moore and H.~Ooguri,
  ``Extremal N=(2,2) 2D Conformal Field Theories and Constraints of Modularity,''
  arXiv:0805.4216 [hep-th].

\bibitem{TMG birth 1} S.~Deser, R.~Jackiw and S.~Templeton, ``Three-Dimensional Massive Gauge Theories,'' Phys.\ Rev.\ Lett.\  {\bf 48}, 975 (1982).

\bibitem{TMG birth 2} S.~Deser, R.~Jackiw and S.~Templeton, ``Topologically massive gauge theories,''
  Annals Phys.\  {\bf 140}, 372 (1982) [Erratum-ibid.\  {\bf 185}, 406.1988\ APNYA,281,409 (1988\ APNYA,281,409-449.2000)].

\bibitem{CTMG deser} S.~Deser, ``Cosmological Topological Supergravity,'' in ``Quantum Theory of Gravity: Essays in honor of the 60th birthday of Bryce S. DeWitt.'' Edited by S.~M.~ Christensen. Published by Adam Hilger Ltd., Bristol, England 1984, p.374.


\bibitem{TMG renormalizable oda} I.~Oda, ``Renormalizability of Topologically Massive Gravity,''
  arXiv:0905.1536 [hep-th].

\bibitem{brown henneaux}  J.~D.~Brown and M.~Henneaux,
  ``Central Charges in the Canonical Realization of Asymptotic Symmetries: An
  Example from Three-Dimensional Gravity,''
  Commun.\ Math.\ Phys.\  {\bf 104}, 207 (1986).

\bibitem{solodhukin HR TMG} S.~N.~Solodukhin, ``Holography with Gravitational Chern-Simons Term,''
  Phys.\ Rev.\  D {\bf 74}, 024015 (2006) [arXiv:hep-th/0509148].

\bibitem{larsen krauss} P.~Kraus and F.~Larsen, ``Holographic gravitational anomalies,'' JHEP {\bf 0601}, 022 (2006)
  [arXiv:hep-th/0508218].

\bibitem{grumiller log}  D.~Grumiller and N.~Johansson, ``Instability in cosmological topologically massive gravity at the chiral
  point,'' JHEP {\bf 0807}, 134 (2008) [arXiv:0805.2610 [hep-th]].

\bibitem{carlip deser photons gravitons} S.~Carlip, S.~Deser, A.~Waldron and D.~K.~Wise, ``Cosmological Topologically Massive Gravitons and Photons,''
  Class.\ Quant.\ Grav.\  {\bf 26}, 075008 (2009) [arXiv:0803.3998 [hep-th]].

\bibitem{AyonBeato:2004fq}
 E.~Ayon-Beato and M.~Hassaine, ``pp waves of conformal gravity with self-interacting source,''
 Annals Phys.\  {\bf 317}, 175 (2005)
 [arXiv:hep-th/0409150].

\bibitem{AyonBeato:2005qq}
 E.~Ayon-Beato and M.~Hassaine, ``Exploring AdS waves via nonminimal coupling,''
 Phys.\ Rev.\  D {\bf 73}, 104001 (2006)
 [arXiv:hep-th/0512074].



\bibitem{valdivia AdS spaces in TMG} M.~Henneaux, C.~Martinez and R.~Troncoso,
  ``Asymptotically anti-de Sitter spacetimes in topologically massive gravity,'' Phys.\ Rev.\  D {\bf 79}, 081502R (2009)
  [arXiv:0901.2874 [hep-th]].

\bibitem{Grumiller consistent bc}
  D.~Grumiller and N.~Johansson, ``Consistent boundary conditions for cosmological topologically massive gravity at the chiral point,''
  Int.\ J.\ Mod.\ Phys.\  D {\bf 17}, 2367 (2009)
  [arXiv:0808.2575 [hep-th]].

\bibitem{LCFT log operators in CFT gurarie}  V.~Gurarie, ``Logarithmic operators in conformal field theory,'' Nucl.\ Phys.\  B {\bf 410}, 535 (1993)
  [arXiv:hep-th/9303160].

\bibitem{LCFT bits and pieces flohr}  M.~Flohr, ``Bits and pieces in logarithmic conformal field theory,''
  Int.\ J.\ Mod.\ Phys.\  A {\bf 18}, 4497 (2003)
  [arXiv:hep-th/0111228].


\bibitem{LCFT algebraic approach gaberdiel} M.~R.~Gaberdiel, ``An algebraic approach to logarithmic conformal field theory,'' Int.\ J.\ Mod.\ Phys.\  A {\bf 18}, 4593 (2003) [arXiv:hep-th/0111260].

\bibitem{GKP} G.~Giribet, M.~Kleban and M.~Porrati, ``Topologically Massive Gravity at the Chiral Point is Not Chiral,''
  JHEP {\bf 0810}, 045 (2008) [arXiv:0807.4703 [hep-th]].


\bibitem{carlip chiral TMG and extremal BF}  S.~Carlip, ``Chiral Topologically Massive Gravity and Extremal B-F Scalars,''
  arXiv:0906.2384 [hep-th].


\bibitem{HR for TMG skenderis taylor} K.~Skenderis, M.~Taylor and B.~C.~van Rees, ``Topologically Massive Gravity and the AdS/CFT Correspondence,''
  arXiv:0906.4926 [hep-th].


\bibitem{simple proof strominger}  A.~Strominger, ``A Simple Proof of the Chiral Gravity Conjecture,'' arXiv:0808.0506 [hep-th].


\bibitem{wald lee symplectic} J.~Lee and R.~M.~Wald, ``Local symmetries and constraints,'' J.\ Math.\ Phys.\  {\bf 31}, 725 (1990).


\bibitem{canonical quantization of TMG buchbinder} I.~L.~Buchbinder, S.~L.~Lyahovich and V.~A.~Krychtin,
  ``Canonical Quantization Of Topologically Massive Gravity,''
  Class.\ Quant.\ Grav.\  {\bf 10}, 2083 (1993).


\bibitem{canonical quantization TMG deser xiang} S.~Deser and X.~Xiang,  ``Canonical Formulations Of Full Nonlinear Topologically Massive Gravity,''
  Phys.\ Lett.\  B {\bf 263}, 39 (1991).

\bibitem{Wald}  R.~M.~Wald, ``Quantum field theory in curved space-time and black hole thermodynamics,''
{\it  Chicago, USA: Univ. Pr. (1994) 205 p}.

\bibitem{BD} N. D. Birrell and P. C. W. Davies, {\it Quantum fields in curved space}, (Cambridge University Press, Cambridge, 1982).

\bibitem{Myung:2008dm}
  Y.~S.~Myung, ``Logarithmic conformal field theory approach to topologically massivegravity,''
  Phys.\ Lett.\  B {\bf 670}, 220 (2008)
  [arXiv:0808.1942 [hep-th]].

\bibitem{ALMMT} A.~Ashtekar, J.~Lewandowski, D.~Marolf, J.~Mourao and T.~Thiemann,
``Quantization of diffeomorphism invariant theories of connections with local
degrees of freedom,''
J.\ Math.\ Phys.\  {\bf 36}, 6456 (1995)
[arXiv:gr-qc/9504018].

\bibitem{where}  D.~Marolf,
  ``Group averaging and refined algebraic quantization: Where are we now?,'' in {\it Proceedings of the
Ninth Marcel-Grossman Conference}, (World Scientific, Singapore, 1994), ed. by V.G. Gurzadyan, R.T. Jantzen, and R. Ruffini; gr-qc/0011112.


\end{thebibliography}
\end{document}